\documentclass[conference]{IEEEtran}
\IEEEoverridecommandlockouts

\usepackage{bm}
\usepackage{dsfont}
\usepackage{stfloats}
\usepackage{relsize}

\usepackage[normalem]{ulem}
\usepackage[dvipsnames]{xcolor}
\usepackage{setspace}

\usepackage[utf8x]{inputenc}
\usepackage[T1]{fontenc}
\usepackage{url}
\usepackage{xfrac}
\usepackage{multicol}
\usepackage{xcolor}
\usepackage{ifthen}
\usepackage{amsfonts}
\usepackage{amssymb}
\usepackage{hyperref}
\usepackage{sansmath}
\usepackage{siunitx}
\usepackage{mathtools}
\usepackage{nccmath}
\usepackage{cite}
\usepackage{oplotsymbl}
\usepackage{systeme}
\usepackage{spalign}
\usepackage{lipsum}
\usepackage{amsmath}
\usepackage{enumitem}
\usepackage{pifont}
\usepackage{subfiles}
\usepackage{accents}
\usepackage{caption}

\usepackage{algpseudocode}
\usepackage{algorithm}
\usepackage{upgreek}

\algnewcommand\algorithmicforeach{\textbf{for each}}
\algdef{S}[FOR]{ForEach}[1]{\algorithmicforeach\ #1\ \algorithmicdo}

\usepackage{eqparbox}
\newdimen{\algindent}
\algnewcommand\LeftComment[2]{%
\hspace{#1\algindent}$\triangleright$ \eqparbox{COMMENT}{#2} \hfill %
}
\algnewcommand\LeftCommentNoTriangle[2]{%
\hspace{#1\algindent} \eqparbox{COMMENT}{#2} \hfill %
}
\algnewcommand\LeftCommentNoIntent[1]{%
$\triangleright$ \eqparbox{COMMENT}{#1} \hfill %
}

\newcommand{\ns}{\hspace{-0.05cm}}

\usepackage{verbatim}
\usepackage{tikz}
\usetikzlibrary{positioning,quotes,calc,fit,shapes,shadows,arrows,trees,mindmap}
\tikzset{block/.style={draw, very thick, minimum height=4cm, align=center}, line/.style={-latex}}
\tikzset{blockV/.style={draw, very thick, text width=2cm, minimum height=2cm, minimum width=4cm, align=center}, line/.style={-latex}}
\tikzset{blockExt/.style={draw, very thick, minimum height=0.7cm, minimum width=0.7cm, align=center}, line/.style={-latex}}
\usepgflibrary{arrows}
\usetikzlibrary{bayesnet}

\definecolor{light-gray}{HTML}{E0E0E0}
\definecolor{carnelian}{rgb}{0.7, 0.11, 0.11}
\definecolor{darkpastelgreen}{rgb}{0.01, 0.75, 0.24}
\definecolor{blue-violet}{rgb}{0.54, 0.17, 0.89}

\makeatletter

\newcommand\notsotiny{\@setfontsize\notsotiny{6.82}{7.5}}
\makeatother

\newcommand*\xbar[1]{%
  \hbox{%
    \vbox{%
      \hrule height 0.5pt 
      \kern0.3ex
      \hbox{%
        \kern-0.05em
        \ensuremath{#1}%
        \kern-0.05em
      }%
    }%
  }%
} 

\newcommand\norm[1]{\left\lVert#1\right\rVert}

\makeatletter
\newcommand{\labeltarget}[1]{\Hy@raisedlink{\hypertarget{#1}{}}}

\usetikzlibrary{fit,calc}

\colorlet{color_pu}{gray!80}
\colorlet{color_pv}{pink!200}
\colorlet{color_uv}{blue!60}
\colorlet{color_phi}{green!60}
\colorlet{color_z}{orange!60}
\definecolor{github-link}{RGB}{0,0,139}

\usepackage{booktabs}
\usepackage{makecell}
\usepackage{multirow}

\usepackage{graphicx}
\usepackage[labelformat=simple]{subcaption}
\usepackage{mwe}
\usepackage[top=1.48cm, bottom=1.5cm,left=0.635in, right=0.635in]{geometry}

\makeatother

\usetikzlibrary{patterns}
\usetikzlibrary{backgrounds}
\usetikzlibrary{decorations.text}

\usepackage{forest}
\usetikzlibrary{trees,positioning,shapes,shadows,arrows.meta}

\newcommand{\appropto}{\mathrel{\vcenter{
  \offinterlineskip\halign{\hfil$##$\cr
    \propto\cr\noalign{\kern2pt}\sim\cr\noalign{\kern-2pt}}}}}

\def\BibTeX{{\rm B\kern-.05em{\sc i\kern-.025em b}\kern-.08em
    T\kern-.1667em\lower.7ex\hbox{E}\kern-.125emX}}
\begin{document}

\title{Time-varying Wireless Channel Tracking\\with Online Parameter Learning via\\ the Birth-Death-Drift Model}

\author{\IEEEauthorblockN{Tiancheng Gao}
\IEEEauthorblockA{\textit{ECE Department} \\
\textit{University of Manitoba}\\
Winnipeg, Canada \\
gaot2@myumanitoba.ca}
\and
\IEEEauthorblockN{Mohamed Akrout}
\IEEEauthorblockA{\textit{EECS Department} \\
\textit{University of Tennessee}\\
Knoxville, USA \\
mohamed.akrout@tennessee.edu}
\and
\IEEEauthorblockN{Faouzi Bellili}
\IEEEauthorblockA{\textit{ECE Department} \\
\textit{University of Manitoba}\\
Winnipeg, Canada \\
faouzi.bellili@umanitoba.ca}
\and
\IEEEauthorblockN{Amine Mezghani}
\IEEEauthorblockA{\textit{ECE Department} \\
\textit{University of Manitoba}\\
Winnipeg, Canada \\
amine.mezghani@umanitoba.ca}
\thanks{This work was supported by the Discovery Grants Program of the Natural Sciences and Engineering Research Council of Canada (NSERC), and the startup funding of University of Tennessee.}
}

\maketitle

\begin{abstract}
Accurate massive MIMO channel state information (CSI) acquisition with low pilot overhead is critical in dynamic propagation environments. Exploiting temporal correlation is key to reducing pilot overhead, yet most existing methods often rely on impractical assumptions. The approximate message passing with side information (AMP-SI) algorithm, built upon a birth-death-drift (BDD) model, represents a significant step in this direction. However, its practical deployment is hindered by three major limitations: reliance on i.i.d. Gaussian sensing matrices, need for perfect BDD parameter knowledge, and a statistically approximate treatment of temporal information. To address these limitations, we introduce BDD-VAMP-EM, a fully automated algorithm that relies on the BDD model, vector AMP (VAMP), and expectation-maximization (EM) in a unified framework. Simulations show that BDD-VAMP-EM consistently outperforms existing benchmarks, particularly under model parameter mismatch, confirming its practical viability.
\end{abstract}

\begin{IEEEkeywords}
Time-varying channel, BDD model, VAMP, EM.
\end{IEEEkeywords}

\section{Introduction}
\label{sec:introduction}

In multiple-input multiple-output (MIMO) and its massive-scale extensions, the performance of physical-layer techniques such as high-order spatial multiplexing \cite{foschini1996layered} and interference suppression \cite{joham2005linear} depend critically on the precision and timeliness of channel state information (CSI). In principle, CSI quality can be improved by transmitting longer training/pilot sequences, especially in presence of a large number of unknown channel coefficients in massive MIMO systems. However, excessive pilot transmission consumes significant time-frequency resources, diminishing throughput for actual data and creating an overhead bottleneck \cite{ngo2013energy}. Conventional solutions that treat each measurement interval independently fail to exploit the inherent temporal correlation \cite{apelfrojd2018channel}, which manifests in two main drawbacks: an inefficient use of spectral resources \cite{bjornson2017massive} and estimation lag and inaccuracy in tracking fast variations \cite{wen2014channel}.

To harness the temporal correlation, a rich body of literature has been developed. Classical methods, such as those based on Kalman filters (KF) \cite{10.1115/1.3662552} or recursive least squares (RLS) \cite{haykin2002adaptive}, are effective only when channel dynamics follow simple, pre-defined models. Their performance deteriorates with non-stationary birth-death dynamics of realistic multi-path components because they lack an inherent mechanism to model structured sparsity evolution \cite{eldar2012compressed}. To address this, modern compressed sensing (CS) techniques, including those based on the approximate message passing (AMP) \cite{donoho2009message}, have proven powerful by leveraging the spatial sparsity in angular domain. In this line of work, the AMP with side information (AMP-SI) algorithm represents a notable leap forward, which integrates the birth-death-drift (BDD) model to exploit temporal correlations across consecutive estimation intervals \cite{ma2019approximate}. To derive the posterior for the current channel state, AMP-SI uses a Bernoulli–Gaussian (BG) prior and forms a Gaussian likelihood by merging the AMP extrinsic message and the latest posterior mean and variance, termed ``side information'' (SI).

Despite its conceptual appeal, AMP-SI also faces certain constraints. Inherited from AMP, the requirement on independent and identically distributed (i.i.d.) Gaussian sensing matrices is often violated in practical systems and potentially leads to algorithmic instability \cite{javanmard2013state}. Also, the performance is contingent upon precise \textit{prior} knowledge of the BDD model parameters: the absence of an online learning mechanism confines its use to idealized settings \cite{ke2020compressive}. Moreover, the treatment of temporal information rests on a Gaussian approximation of the entire posterior parameterized solely by the mean and variance, which discards richer statistical structures, especially in low signal-to-noise ratio (SNR) regimes or when the sparsity pattern is uncertain \cite{vila2013expectation}.

This paper proposes a fully automated algorithm for joint channel tracking and parameter learning, coined BDD-VAMP-EM, to address the three core limitations of AMP-SI:
\vspace{-0.05cm}
\begin{itemize}[leftmargin=*]
    \item \textit{Robustness to practical sensing matrices}: we build upon vector AMP (VAMP) \cite{rangan2019vector} to provide reliable convergence under broader conditions on sensing matrices than AMP.
    
    \item \textit{Autonomy in dynamic environments}: we seamlessly integrate an expectation-maximization (EM) \cite{bishop2006pattern} procedure within the message-passing iteration loop to enable online, joint estimation of the BDD parameters and the channel state.
    
    \item \textit{Statistically faithful temporal fusion}: we assume the posterior is a Bernoulli Gaussian-mixture (BGM) distribution and propagate the hyper-parameters as SI to allow for a richer statistical structure than moment-based approximations.
\end{itemize}

The remainder of this paper is structured as follows. Section \ref{sec:preliminary} details the system model and formulates the channel tracking problem. Section \ref{sec:BDD-VAMP-EM} presents the derivation of the proposed BDD-VAMP-EM algorithm. Section \ref{sec:numerical-results} provides a comprehensive performance evaluation against state-of-the-art benchmarks. Finally, Section \ref{sec:conclusion} concludes the paper.

\textbf{Notations}: We use Sans Serif fonts (e.g., $\mathsf{x}$) for random variables and Serif fonts (e.g., $x$) for realizations. We use boldface lowercase and uppercase letters for vectors (e.g., $\bm{\mathsf{x}}$ and $\bm{x}$) and matrices (e.g., $\bm{\mathsf{X}}$ and $\bm{X}$), respectively. Vectors are column-wise by default. $\bm{X}^{\textsf{T}}$ and $\bm{X}^{\textsf{H}}$ refer to the transpose and hermitian of matrix $\bm{X}$, respectively. $\mathbf{I}_{N}$ denotes the $N \times N$ identity matrix and $\mathrm{vec}(\bm{X})$ denotes the vectorization operation that stacks the columns of $\bm{X}$ into a vector.
Moreover, $\mathrm{Pr}\{\cdot\}$ denotes the probability of an event, while $p_{\mathsf{x}}(x;\bm{\theta})$, $p_{\bm{\mathsf{x}}} (\bm{x};\bm{\theta})$, and $p_{\bm{\mathsf{X}}} (\bm{X};\bm{\theta})$ denote the probability density function (PDF) of random variables/vectors/matrices, as being parameterized by $\bm{\theta}$. $\mathcal{CN}(\bm{x};\widehat{\bm{x}},\boldsymbol{\Sigma})$ refers to the multivariate Gaussian distribution of a complex-valued vector $\bm{x}\in\mathbb{C}^N$ and $\delta(\boldsymbol{x})$ refers to the multi-dimensional Dirac delta distribution. In addition, the short-hand notation $\sim$ stands for ``distributed according to''. The symbol $\otimes$ denotes the Kronecker product operator. We also use $\int$ as a shorthand for $\int_{-\infty}^{\infty}$.

\section{System Model and Problem Formulation}
\label{sec:preliminary}

Consider a point-to-point down-link communication system comprising a base station (BS) and a mobile user equipment (UE), equipped with $N_{\textsf{Tx}}$ and $N_{\textsf{Rx}}$ antenna elements, respectively, each forming a one-dimensional uniform linear array (ULA). We consider block fading scenario, where the channel matrix, denoted by $\bm{H}_t \in \mathbb{C}^{N_{\textsf{Rx}} \times {N}_{\textsf{Tx}}}$ at each time step $t \in \{1,2,\ldots,T\}$, remains constant over each $t^\text{th}$ coherence block but may change in subsequent blocks. The physical channel $\bm{H}_t$ can be represented in angular domain through the virtual channel model \cite{sayeed2002deconstructing} as
\begin{equation}
\label{eq:vir-chnl_H}
    \bm{H}_t = \bm{\mathcal{A}}_\textsf{Rx}\, \bm{H}_{\textsf{v},t}\, \bm{\mathcal{A}}_\textsf{Tx}^\textsf{H},
\end{equation}

\noindent where $\bm{H}_{\textsf{v},t}$ is the virtual channel matrix and $\bm{\mathcal{A}}_{\textsf{Rx}} \in \mathbb{C}^{N_{\textsf{Rx}} \times {N}_{\textsf{Rx}}}$ and $\bm{\mathcal{A}}_\textsf{Tx} \in \mathbb{C}^{N_{\textsf{Tx}} \times {N}_{\textsf{Tx}}}$ are the normalized discrete Fourier transform (DFT) matrices.

The virtual channel $\bm{H}_{\textsf{v},t}$ in massive MIMO systems often exhibits a sparse structure due to the limited number of dominant scattering clusters in the propagation environment. To capitalize on this, we model the dynamics of each element in $\bm{H}_{\textsf{v},t}$ using a BDD process, which tracks the appearance, disappearance, and gradual variation of significant channel components over time. Let $h_{\textsf{v},t}$ denote a generic entry in $\bm{H}_{\textsf{v},t}$. The evolution of $h_{\textsf{v},t}$ in time given $\mathsf{h}_{\textsf{v},t-1}$ involves the following four cases in the BDD model:
\begin{itemize}
    \item[$i)$] \underline{\textit{Birth Case}}: zero entry becomes nonzero, i.e., $h_{\textsf{v},t-1}=0$ and $\mathsf{h}_{\textsf{v},t} \sim \mathcal{CN}(h_{\textsf{v},t}; 0, \eta_{t}^{-1})$, characterized by the conditional probability $\mathrm{Pr}\{ \mathsf{h}_{\textsf{v},t}\neq 0 | \mathsf{h}_{\textsf{v},t-1}=0 \} \triangleq \beta_{\textsf{B},t}$. 
    
    \item[$ii)$] \underline{\textit{Death Case}}: nonzero entry becomes zero, i.e., $\mathsf{h}_{\textsf{v},t-1} \sim \mathcal{CN}(h_{\textsf{v},t-1}; 0, \eta_{t-1}^{-1})$ and $h_{\textsf{v},t}=0$, characterized by the conditional probability $\mathrm{Pr}\{ \mathsf{h}_{\textsf{v},t}=0 | \mathsf{h}_{\textsf{v},t-1}\neq 0 \} \triangleq \beta_{\textsf{D},t}$. 
    
    \item[$iii)$] \underline{\textit{Drift Case}}: nonzero entry remains nonzero, i.e., $\mathsf{h}_{\textsf{v},t-1} \sim \mathcal{CN}(h_{\textsf{v},t-1}; 0, \eta_{t-1}^{-1})$ and $h_{\textsf{v},t} = \alpha_t\, h_{\textsf{v},t-1} + \sqrt{1-\alpha_t^2}\, n_t$ with $\mathsf{n}_t \sim \mathcal{CN}(n_t; 0, \eta_{t}^{-1})$, characterized by the conditional probability $\mathrm{Pr}\{ \mathsf{h}_{\textsf{v},t}\neq 0 | \mathsf{h}_{\textsf{v},t-1}\neq 0 \} = 1-\beta_{\textsf{D},t}$. Here, we assume $\eta_t \approx \eta_{t-1}$ and resort to the scaling factor $\sqrt{1-\alpha_t^2}$ to maintain stationarity and ensure the power of non-zero entries to remain consistent.

    \item[$iv)$] \underline{\textit{Zero Case}}: zero entry remains zero, i.e., $h_{\textsf{v},t-1}=0$ and $h_{\textsf{v},t}=0$, characterized by the conditional probability $\mathrm{Pr}\{ \mathsf{h}_{\textsf{v},t}=0 | \mathsf{h}_{\textsf{v},t-1}=0 \} = 1-\beta_{\textsf{B},t}$. 
\end{itemize}
\vspace{0.1cm}

Over each $t^\text{th}$ coherence block, the received signal at the UE can be expressed as
\begin{equation}
\label{eq:SS-model_pilot-observation}
    \bm{Y}_t = \bm{H}_t\, \bm{P}_t + \bm{W}_t,
\end{equation}

\noindent where, $\bm{P}_t = [\bm{p}_{t,1}, \ldots, \bm{p}_{t,N_{\textsf{P}}}] \in \mathbb{C}^{N_{\textsf{Tx}}\times {N}_{\textsf{P}}}$ contains the ${N}_{\textsf{P}}$ uncorrelated complex-valued pilot vectors transmitted over coherence block $t$, and $\bm{Y}_t = [\bm{y}_{t,1}, \ldots, \bm{y}_{t,N_{\textsf{P}}}] \in \mathbb{C}^{N_{\textsf{Rx}}\times {N}_{\textsf{P}}}$ concatenates the associated received signals. The measurement noise $\bm{W}_t \in \mathbb{C}^{N_{\textsf{Rx}}\times {N}_{\textsf{P}}}$ is an i.i.d. additive white Gaussian noise (AWGN) matrix with each element following $\mathcal{CN}(w_t; 0, \gamma_{w}^{-1})$. Plugging (\ref{eq:vir-chnl_H}) into (\ref{eq:SS-model_pilot-observation}) and vectorizing yields
\begin{equation}
\label{eq:SS-model_measurement-eq}
\begin{aligned}[b]
    \bm{y}_t &\triangleq \mathrm{vec}(\bm{Y}_t)\\
    &= \mathrm{vec}\big( \bm{\mathcal{A}}_\textsf{Rx}\, \bm{H}_{\textsf{v},t}\, \bm{\mathcal{A}}_\textsf{Tx}^\textsf{H}\, \bm{P}_t \big) + \mathrm{vec}(\bm{W}_t)\\
    &= \underbrace{\Big(\big(\bm{\mathcal{A}}_\textsf{Tx}^\textsf{H}\, \bm{P}_t\big)^\textsf{T} \otimes \bm{\mathcal{A}}_\textsf{Rx}\Big)}_{\bm{A}_t} \underbrace{\mathrm{vec}\big(\bm{H}_{\textsf{v},t}\big)}_{\bm{x}_t} + \underbrace{\mathrm{vec}(\bm{W}_t)}_{\bm{w}_t},
\end{aligned}
\end{equation}

\noindent which recasts the channel estimation problem in (\ref{eq:SS-model_pilot-observation}) as a linear inverse problem. Estimating $\bm{x}_t \in \mathbb{C}^{N_x}$ from the observed $\bm{y}_t \in \mathbb{C}^{N_y}$ is equivalent to estimating the channel matrix $\bm{H}_t \in \mathbb{C}^{Ny\times N_x}$, with $N_x = N_{\textsf{Rx}}\, N_{\textsf{Tx}}$ and $N_y = N_{\textsf{Rx}}\, {N}_{\textsf{P}}$ Furthermore, the temporal evolution of $\bm{x}_t = \mathrm{vec}\big( \bm{H}_{\textsf{v},t} \big)$ is a BDD process and can be compactly represented by an element-wise random function $\bm{f}_{\textsf{BDD}}(\cdot)$ as follows:
\begin{equation}
\label{eq:SS-model_process-eq}
    \bm{x}_t = \bm{f}_{\textsf{BDD}}(\bm{x}_{t-1}, \bm{\theta}_t),
\end{equation}

\noindent with $\bm{\theta}_t = [\beta_{\textsf{B},t}, \beta_{\textsf{D},t}, \alpha_t, \eta_t]^\textsf{T}$ being the set of parameters that characterize the BDD process at time step $t$. As a result, the state-space model of the system is defined by (\ref{eq:SS-model_measurement-eq}) and (\ref{eq:SS-model_process-eq}). Note that the algorithm developed in this paper tracks the channel matrix as it evolves across successive coherence blocks. At each time step $t$, it updates the channel matrix associated with the $t^{\text{th}}$ coherence block. The time index $t \in \{1,2,\ldots,T\}$ should not be confused with the time index used for pilot transmissions within each block, which ranges from $1$ to $N_{\textsf{P}}$.

\section{The BDD-VAMP-EM algorithm}
\label{sec:BDD-VAMP-EM}

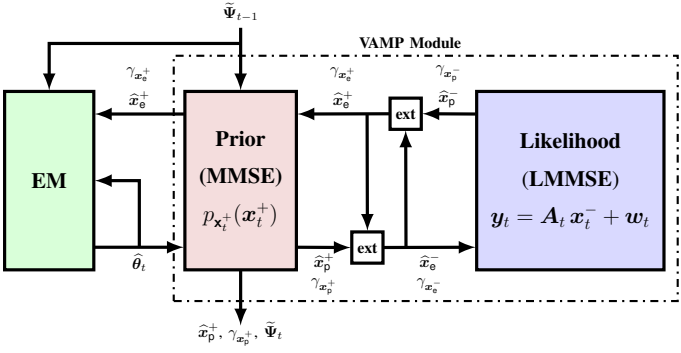
\begin{figure}[!h]
    \centering
    \begin{tikzpicture}[thick,scale=0.59, every node/.style={transform shape}]
    \node[block, fill=carnelian!15, minimum width=2.5cm] (blk_pri) {\Large{\textbf{Prior}} \\\\ \Large{\textbf{(MMSE)}} \\\\ \Large{$p_{\bm{\mathsf{x}}_{t}^{+}}(\bm{x}_{t}^{+})$}};
    
    \node[block, fill=blue!15, right=4cm of blk_pri, minimum width=4.2cm] (blk_lik) {\Large{\textbf{Likelihood}} \\\\ \Large{(\textbf{LMMSE})} \\\\ \Large{$\bm{y}_{t} = \bm{A}_{t}\, \bm{x}_{t}^{-}+\bm{w}_t$}};
    
    \node[block, fill=green!15, left=2cm of blk_pri, minimum width=2cm] (blk_EM) {\Large{\textbf{EM}}};

    \node[blockExt,right=of blk_pri, xshift=0.2cm, yshift=-1.5cm] (ext_pri2lik) {$\bm{\textbf{ext}}$};
    \node[blockExt,left=of blk_lik, xshift=-0.2cm, yshift=1.5cm] (ext_lik2pri) {$\bm{\textbf{ext}}$};

    \draw [-latex,very thick] ([yshift=-4.25em] blk_pri.east) -- node [midway,below=0em,align=center] {$\widehat{\bm{x}}_{\textsf{p}}^+$} node [midway,below=1.6em,align=center] {$\gamma_{\bm{x}_{\textsf{p}}^+}$} (ext_pri2lik.west);
    \draw [-latex,very thick] (ext_pri2lik) -- node [midway,below=0em,align=center] {$\widehat{\bm{x}}_{\textsf{e}}^{-}$} node [midway,below=1.6em,align=center] {$\gamma_{\bm{x}_{\textsf{e}}^{-}}$} ([yshift=-4.25em] blk_lik.west) node [pos=0.24] (restPoint_extArrow_rise) {};
    
    \draw [-latex,very thick] ([yshift=4.25em] blk_lik.west) -- node [midway,above=0em,align=center] {$\widehat{\bm{x}}_{\textsf{p}}^-$} node [midway,above=1.7em,align=center] {$\gamma_{\bm{x}_{\textsf{p}}^-}$} (ext_lik2pri.east);
    \draw [-latex,very thick] (ext_lik2pri) -- node [midway,above=0em,align=center] {$\widehat{\bm{x}}_{\textsf{e}}^{+}$} node [midway,above=1.7em,align=center] {$\gamma_{\bm{x}_{\textsf{e}}^{+}}$} ([yshift=4.25em]blk_pri.east) node [pos=0.24] (restPoint_extArrow_drop) {};

    \draw [-latex,very thick] ([yshift=4.25em] blk_pri.west) -- node [midway,above=0em,align=center] {$\widehat{\bm{x}}_{\textsf{e}}^{+}$} node [midway,above=1.7em,align=center] {$\gamma_{\bm{x}_{\textsf{e}}^{+}}$} ([yshift=4.25em] blk_EM.east);
    
    \draw [-latex,very thick] ([yshift=-4.25em] blk_EM.east) -- node [midway,below=0em,align=center] {$\widehat{\bm{\theta}}_t$} ([yshift=-4.25em] blk_pri.west) node [pos=0.5] (restPoint_EMArrow_feedback) {};

    \draw [-latex,very thick] (restPoint_extArrow_rise.center) -- (ext_lik2pri.south);
    \draw [-latex,very thick] (restPoint_extArrow_drop.center) -- (ext_pri2lik.north);

    \draw [-latex,very thick] ([yshift=4em] blk_pri.north) -- (blk_pri.north) node [pos=0.25] (restPoint_inputArrow) {};
    \draw [-latex,very thick] (restPoint_inputArrow.center) -- ([yshift=3em] blk_EM.north) -- (blk_EM.north);
    \node [] at ([yshift=5em] blk_pri.north) {$\widetilde{\bm{\Psi}}_{t-1}$};
    
    \draw [-latex,very thick] (blk_pri.south) -- ([yshift=-3.5em] blk_pri.south);
    \node [] at ([yshift=-4em] blk_pri.south) {$\widehat{\bm{x}}_{\textsf{p}}^+$, $\gamma_{\bm{x}_{\textsf{p}}^+}$, $\widetilde{\bm{\Psi}}_t$};

    \draw [-latex,very thick] (restPoint_EMArrow_feedback.center) -- ([yshift=4.25em] restPoint_EMArrow_feedback.center) -- (blk_EM.east);

    \draw[dashdotted] (-1.5,-2.7) rectangle +(11.3,5.5);
    \node[] at (3.8,3.1) {$\textbf{\textrm{VAMP Module}}$};
    
\end{tikzpicture}
    \vspace{-0.5cm}
    \caption{Block diagram of BDD-VAMP-EM at $t > 0$. The algorithm iterates between the EM module and the VAMP module (itself consisting of an MMSE block and an LMMSE block). SI is in the form of posterior hyper-parameters $\widetilde{\bm{\Psi}}_{t-1}$ from step $t-1$. In each internal iteration within the time step, the EM module updates the parameter estimate of the prior distribution, and the compound VAMP module executes one iteration of standard VAMP to recover $\bm{x}_t$. Upon reaching the maximum number of iterations, the MMSE block returns the estimate $\widehat{\bm{x}}_{\textsf{p}}^+$ and precision $\gamma_{\bm{x}_{\textsf{p}}^+}$, and passes $\widetilde{\bm{\Psi}}_t$ as SI to $t+1$.}
    \label{fig:BDD-VAMP-EM_block-diagram}
\end{figure}

This section presents the derivation of the BDD-VAMP-EM algorithm based on the state-space model in (\ref{eq:SS-model_measurement-eq}) and (\ref{eq:SS-model_process-eq}). The algorithm jointly tracks $\bm{x}_t$ and updates the parameter vector $\bm{\theta}_t$ at every time step $t$. For clarity, Fig.~\ref{fig:BDD-VAMP-EM_block-diagram} provides a block diagram illustrating the algorithm’s structure, in which the posterior distribution’s hyper-parameters $\widetilde{\bm{\Psi}}_t$ are explicitly computed and used as SI for $t+1$. As we show, under an i.i.d. BG prior for the initial state $\bm{x}_0$, the posterior distribution of $\bm{x}_t$ at any $t>0$ becomes element-wise independent and exhibits a BGM structure with $\widetilde{\bm{\Psi}}_t$ collectively representing the weights, means, and precisions of these Dirac-delta and Gaussian components.

In this section, $\bm{\theta}$ and $\widehat{\bm{\theta}}$ denote the vector of unknown parameters and its EM estimate, respectively. In addition, $\rho$, $\mu$, and $\gamma$ denote the prior hyper-parameters, while $\widetilde{\rho}$, $\widetilde{\mu}$, and $\widetilde{\gamma}$ denote their posterior counterparts. In the VAMP module, $\bm{x}$ is split into two auxiliary variables $\bm{x}^+$ and $\bm{x}^-$, and posterior and extrinsic variables are distinguished using the subscripts ``$\textsf{p}$'' and ``$\textsf{e}$'', respectively. Regarding the iterative processes, $t$ indexes the discrete time steps of block-fading transmissions, while $k$ indexes the internal iterations of the VAMP and EM modules. For clarity, variable subscripts are arranged in the following order: element index, variable type, time step $t$, and iteration number $k$. For example, $\widehat{x}_{n,\textsf{e},t,k}$ denotes the $n^\text{th}$ entry of the extrinsic estimate $\widehat{\bm{x}}_{\textsf{e}}$ at time step $t$ and iteration $k$.

In the sequel, we start by establishing the recursive relationship between the BGM prior and posterior across successive time steps. Then we derive the parameter estimation within the EM module. The prior at $t=0$ is assumed to be a BG distribution, which is a special case consistent with the general formulation.

\subsection{Prior and Posterior Quantities Evolution}
\label{subsec:BDD-VAMP-EM_t>0_pri-pst-evo-with-MMSE}

We begin by assuming that the element-wise posterior at time step $t-1$ and final iteration $k=K$ is a BGM distribution with $M_{t-1}$ Gaussian components. For notational simplicity, we omit the common conditioning variables $\mathsf{x}_{n,\textsf{e},t-1,K}^+$ and $\widehat{x}_{n,\textsf{e},t-1,K}^+$ in all density functions, and suppress the explicit time index $t$. We also replace $t-1$ with the subscript ``$\textrm{old}$''. Accordingly, the posterior $p_{\mathsf{x}_{n,t-1}^+ \mid \mathsf{x}_{n,\textsf{e},t-1}^+}\big( x_{t-1} \big| \widehat{x}_{n,\textsf{e},t-1,K}^+; \widetilde{\bm{\Psi}}_{n,t-1} \big)$ simplifies to
\begin{equation}
\label{eq:BDD-VAMP-EM_pst-xold}
\begin{aligned}[b]
    &p_{\mathsf{x}_{n,\textrm{old}}^+}(x_{\textrm{old}}; \widetilde{\bm{\Psi}}_{n,\textrm{old}})\\
    &\hspace{0cm}= \widetilde{\rho}_{n,\delta,\textrm{old},K}\, \delta(x_{\textrm{old}})\\
    &\hspace{0.3cm}+\ns\ns \sum_{m=1}^{M_{\textrm{old}}} \widetilde{\rho}_{n,{{\scriptscriptstyle\mathcal{N}}_{\ns{m}}},\textrm{old},K}\, \mathcal{CN}\big( x_{\textrm{old}}; \widetilde{\mu}_{n,{{\scriptscriptstyle\mathcal{N}}_{\ns{m}}},\textrm{old},K}, \widetilde{\gamma}_{n,{{\scriptscriptstyle\mathcal{N}}_{\ns{m}}},\textrm{old},K}^{-1} \big),
\end{aligned}
\end{equation}

\noindent Here, the subscript ${\scriptstyle\mathcal{N}}_{\ns{m}}$ denotes the $m^\text{th}$ Gaussian component in the BGM distribution. The parameters $\widetilde{\rho}_{n,\delta,\textrm{old},K}$, $\widetilde{\rho}_{n,{{\scriptscriptstyle\mathcal{N}}_{\ns{m}}},\textrm{old},K}$, $\widetilde{\mu}_{n,{{\scriptscriptstyle\mathcal{N}}_{\ns{m}}},\textrm{old},K}$, and $\widetilde{\gamma}_{n,{{\scriptscriptstyle\mathcal{N}}_{\ns{m}}},\textrm{old},K}$ for all elements $n$ and Gaussian components $m$ are assembled into the matrix $\widetilde{\bm{\Psi}}_{\textrm{old}}$ and passed to time step $t$ as SI. The column $\widetilde{\bm{\Psi}}_{n,\textrm{old}}$ contains the parameters for element $n$. Evolving from the posterior for $\bm{x}_{\textrm{old}}$ in (\ref{eq:BDD-VAMP-EM_pst-xold}), the prior for $\bm{x}$ inherits the element-wise separable structure and introduces one additional Gaussian component (i.e., $M=M_{\textrm{old}}+1$):
\begin{equation}
\label{eq:BDD-VAMP-EM_pri-xt}
    p_{\mathsf{x}_n^+}(x; \bm{\Psi}_n) \ns=\ns \rho_{n,\delta}\, \delta(x) + \ns\ns\sum_{m=1}^M \ns \rho_{n,{{\scriptscriptstyle\mathcal{N}}_{\ns{m}}}}\, \mathcal{CN}(x; \mu_{n,{{\scriptscriptstyle\mathcal{N}}_{\ns{m}}}}, \gamma_{n,{{\scriptscriptstyle\mathcal{N}}_{\ns{m}}}}^{-1}).
\end{equation}

\noindent Let $\bm{\theta} = [\beta_{\textsf{B}}, \beta_{\textsf{D}}, \alpha, \eta]^\textsf{T}$ denote the BDD model parameters at $t$. The update rules of the prior hyper-parameters $\bm{\Psi}_n$ in (\ref{eq:BDD-VAMP-EM_pri-xt}) as a function of $\widetilde{\bm{\Psi}}_{n,\textrm{old}}$ in (\ref{eq:BDD-VAMP-EM_pst-xold}) and $\bm{\theta}$ are specified explicitly as
\begin{equation}
\label{eq:BDD-VAMP-EM_pri-xt_parameters}
\begin{aligned}[b]
    \rho_{n,\delta} &= \widetilde{\rho}_{n,\delta,\textrm{old},K}\, (1-\beta_{\textsf{B}}) + \big(1-\widetilde{\rho}_{n,\delta,\textrm{old},K}\big)\, \beta_{\textsf{D}},\\
    \rho_{n,{{\scriptscriptstyle\mathcal{N}}_{\ns{m}}}} &= 
    \begin{cases}
      \widetilde{\rho}_{n,{{\scriptscriptstyle\mathcal{N}}_{\ns{m}}},\textrm{old},K}\, (1-\beta_{\textsf{D}})& \textrm{if } 1\leq m\leq M_{\textrm{old}},\\
      \widetilde{\rho}_{n,\delta,\textrm{old},K}\, \beta_{\textsf{B}}& \textrm{if } m = M,
    \end{cases}\\
    \mu_{n,{{\scriptscriptstyle\mathcal{N}}_{\ns{m}}}} &= 
    \begin{cases}
      \alpha\, \widetilde{\mu}_{n,{{\scriptscriptstyle\mathcal{N}}_{\ns{m}}},\textrm{old},K}& \textrm{if } 1\leq m\leq M_{\textrm{old}},\\
      0& \textrm{if } m = M,
    \end{cases}\\
    \gamma_{n,{{\scriptscriptstyle\mathcal{N}}_{\ns{m}}}} &= 
    \begin{cases}
      \big(1-\alpha^2\big)\, \eta^{-1} \ns+ \alpha^2\, \widetilde{\gamma}_{n,{{\scriptscriptstyle\mathcal{N}}_{\ns{m}}},\textrm{old},K}^{-1}& \textrm{if } 1\leq m\leq M_{\textrm{old}},\\
      \eta& \textrm{if } m = M.
    \end{cases}
\end{aligned}
\end{equation}

\noindent The posterior for $\bm{x}$ at iteration $k$ given the VAMP extrinsic message $\mathcal{CN}\big( \widehat{\bm{x}}_{\textsf{e},k}^+; \bm{x}, \gamma_{\bm{x}_{\textsf{e}}^+\ns,k}^{-1} \mathbf{I}_N \big)$ is therefore element-wise independent and can be expressed as
\begin{equation}
\label{eq:BDD-VAMP-EM_pst-xt}
\begin{aligned}[b]
    &p_{\mathsf{x}_n^+ \mid \mathsf{x}_{n,\textsf{e}}^+}\big( x \big| \widehat{x}_{n,\textsf{e},k}^+; \bm{\Psi}_n, \gamma_{\bm{x}_{\textsf{e}}^+\ns,k} \big)\\ 
    &\hspace{0.1cm}= \widetilde{\rho}_{n,\delta,k}\, \delta(x) +\ns\ns \sum_{m=1}^M \widetilde{\rho}_{n,{\scriptscriptstyle\mathcal{N}}_{\ns{m}},k}\, \mathcal{CN}\big(x; \widetilde{\mu}_{n,{\scriptscriptstyle\mathcal{N}}_{\ns{m}},k}, \widetilde{\gamma}_{n,{\scriptscriptstyle\mathcal{N}}_{\ns{m}},k}^{-1}\big),
\end{aligned}
\end{equation}

\noindent with the hyper-parameters given by
\begin{equation}
\label{eq:BDD-VAMP-EM_pst-xt_hyperparms}
\begin{aligned}[b]
    \widetilde{\rho}_{n,\delta,k} &= \frac{ \rho_{n,\delta}\, \mathcal{CN}\big(\widehat{x}_{n,\textsf{e},k}^+; 0, \gamma_{\bm{x}_{\textsf{e}}^+\ns,k}^{-1}\big) }{ p_{\mathsf{x}_{n,\textsf{e}}^+}\big( \widehat{x}_{n,\textsf{e},k}^+; \bm{\Psi}_n, \gamma_{\bm{x}_{\textsf{e}}^+\ns,k} \big) },\\
    \widetilde{\rho}_{n,{\scriptscriptstyle\mathcal{N}}_{\ns{m}},k} &= \frac{ \rho_{n,{\scriptscriptstyle\mathcal{N}}_{\ns{m}}}\, \mathcal{CN}\big(\widehat{x}_{n,\textsf{e},k}^+; \mu_{n,{\scriptscriptstyle\mathcal{N}}_{\ns{m}}}, \gamma_{n,{\scriptscriptstyle\mathcal{N}}_{\ns{m}}}^{-1} \ns\ns + \ns \gamma_{\bm{x}_{\textsf{e}}^+\ns,k}^{-1}\big) }{ p_{\mathsf{x}_{n,\textsf{e}}^+}\big( \widehat{x}_{n,\textsf{e},k}^+ ; \bm{\Psi}_n, \gamma_{\bm{x}_{\textsf{e}}^+\ns,k} \big) },\\
    \widetilde{\gamma}_{n,{\scriptscriptstyle\mathcal{N}}_{\ns{m}},k} &= \gamma_{n,{\scriptscriptstyle\mathcal{N}}_{\ns{m}}} + \gamma_{\bm{x}_{\textsf{e}}^+\ns,k},\\
    \widetilde{\mu}_{n,{\scriptscriptstyle\mathcal{N}}_{\ns{m}},k} &= \widetilde{\gamma}_{n,{\scriptscriptstyle\mathcal{N}}_{\ns{m}},k}^{-1} \big( \gamma_{n,{\scriptscriptstyle\mathcal{N}}_{\ns{m}}}\, \mu_{n,{\scriptscriptstyle\mathcal{N}}_{\ns{m}}} \ns\ns + \ns \gamma_{\bm{x}_{\textsf{e}}^+\ns,k}\, \widehat{x}_{\textsf{e},k}^+ \big).
\end{aligned}
\end{equation}

\noindent The normalization factor $p_{\mathsf{x}_{n,\textsf{e}}^+}\big( \widehat{x}_{n,\textsf{e},k}^+; \bm{\Psi}_n, \gamma_{\bm{x}_{\textsf{e}}^+\ns,k} \big)$ makes (\ref{eq:BDD-VAMP-EM_pst-xt}) a valid PDF by ensuring $\widetilde{\rho}_{n,\delta,k} + \sum_{m=1}^M \widetilde{\rho}_{n,{\scriptscriptstyle\mathcal{N}}_{\ns{m}},k} = 1$. Notably, the posterior (\ref{eq:BDD-VAMP-EM_pst-xt}) retains the BGM structure as the prior (\ref{eq:BDD-VAMP-EM_pri-xt}) with updated hyper-parameters. Upon reaching the maximum number of iterations $K$, the posterior hyper-parameters for element $n$ are assembled into $\widetilde{\bm{\Psi}}_n$, allowing the final posterior to be denoted compactly as $p_{\mathsf{x}_n \mid \mathsf{x}_{n,\textsf{e}}^+}\big( x \big| \widehat{x}_{n,\textsf{e},K}^+; \widetilde{\bm{\Psi}}_n \big)$.

\subsection{EM Parameter Estimation}
\label{subsec:BDD-VAMP-EM_t>0_EM}

In the EM module, let $\widehat{\bm{\theta}}_{k} = [\widehat{\beta}_{\textsf{B},k}, \widehat{\beta}_{\textsf{D},k}, \widehat{\alpha}_{k}, \widehat{\eta}_{k}]^\textsf{T}$ denote the EM estimate from iteration $k$ corresponding to the unknown BDD parameters in the prior (\ref{eq:BDD-VAMP-EM_pri-xt}) at $t$. In line with the VAMP principle, we treat the extrinsic estimate $\widehat{\bm{x}}_{\textsf{e},k}^+$ as a pseudo-observation of $\bm{x}$ corrupted by i.i.d. AWGN and introduce a latent variable $\bm{z}$ with elements $z_n\in \{0,1,\ldots,M\}$ for $n=1,\ldots,N_x$ to indicate the component of origin for $\widehat{x}_{n,\textsf{e},k}^+$, i.e., $z_n = 0$ corresponds to $x_n = 0$ and $z_n = m$ ($1\leq{m}\leq{M}$) corresponds to the $m^\text{th}$ Gaussian component. 

In the E-step, the expected complete-data log-likelihood $Q_{k+1}\big(\bm{\theta}, \widehat{\bm{\theta}}_{k}\big)$ can be expressed as
\begin{equation}
\label{eq:BDD-VAMP-EM_EM@t_avg-complete-log-lik}
    Q_{k+1}\big(\bm{\theta}, \widehat{\bm{\theta}}_{k}\big) = q_{\beta,k+1}(\beta_{\textsf{B}},\beta_{\textsf{D}}) - q_{(\alpha,\eta),k+1}(\alpha, \eta) + \mathrm{const},
\end{equation}

\noindent with
\begin{equation*}
\begin{aligned}
    &q_{\beta,k+1}(\beta_{\textsf{B}},\beta_{\textsf{D}})\nonumber\\
    &= \ns \sum_{n=1}^{N_x} \ln \ns \Big(\widetilde{\rho}_{n,\delta,\textrm{old}}\, (1-\beta_{\textsf{B}}) + \big(1-\widetilde{\rho}_{n,\delta,\textrm{old}}\big)\, \beta_{\textsf{D}}\Big) \mathcal{P}_{n,\delta,k+1}\nonumber\\
    &\hspace{0.4cm}+ \ln{(1-\beta_{\textsf{D}})} \sum_{n=1}^N \ns \sum_{m=1}^{M_{\textrm{old}}} \ns \mathcal{P}_{n,{\scriptscriptstyle\mathcal{N}}_{\ns{m}},k+1} \ns + \ln{\beta_\textsf{B}} \sum_{n=1}^N \ns \mathcal{P}_{n,{\scriptscriptstyle\mathcal{N}}_{\ns{M}}\ns,k+1}
\end{aligned}
\end{equation*}

\noindent and
\begin{equation*}
\begin{aligned}[b]
    &q_{(\alpha,\eta),k+1}(\alpha, \eta)\nonumber\\
    &= \ns \sum_{n=1}^{N_x} \hspace{-0.05cm} \sum_{m=1}^{M_{\textrm{old}}} \hspace{-0.05cm} \ln \ns \Big( \ns \big(1-\alpha^2\big) \eta^{-1} \ns + \alpha^2\, \widetilde{\gamma}_{n,{\scriptscriptstyle\mathcal{N}}_{\ns{m}},\textrm{old}}^{-1} + \gamma_{\bm{x}_{\textsf{e}}^+\ns,k}^{-1}\Big) \mathcal{P}_{n,{\scriptscriptstyle\mathcal{N}}_{\ns{m}},k+1}\nonumber\\
    &\hspace{0.4cm}+ \ns \sum_{n=1}^{N_x} \ns \sum_{m=1}^{M_{\textrm{old}}} \ns \frac{\big|\widehat{x}_{n,\textsf{e},k}^+ - \alpha\, \widetilde{\mu}_{n,{\scriptscriptstyle\mathcal{N}}_{\ns{m}},\textrm{old}}\big|^2}{\big(1-\alpha^2\big) \eta^{-1} \ns + \alpha^2\, \widetilde{\gamma}_{n,{\scriptscriptstyle\mathcal{N}}_{\ns{m}},\textrm{old}}^{-1} + \gamma_{\bm{x}_{\textsf{e}}^+\ns,k}^{-1}}\, \mathcal{P}_{n,{\scriptscriptstyle\mathcal{N}}_{\ns{m}},k+1}\nonumber\\
    &\hspace{0.8cm}+ \ln\big(\eta^{-1} \ns + \gamma_{\bm{x}_{\textsf{e}}^+\ns,k}^{-1}\big) \sum_{n=1}^{N_x} \mathcal{P}_{n,{\scriptscriptstyle\mathcal{N}}_{\ns{M}}\ns,k+1}\nonumber\\
    &\hspace{1.2cm}+ \frac{1}{\eta^{-1} \ns + \gamma_{\bm{x}_{\textsf{e}}^+\ns,k}^{-1}} \sum_{n=1}^{N_x} \big|\widehat{x}_{n,\textsf{e},k}^+\big|^2\, \mathcal{P}_{n,{\scriptscriptstyle\mathcal{N}}_{\ns{M}}\ns,k+1} + \mathrm{const}.
\end{aligned}
\end{equation*}

\noindent Here, $\mathcal{P}_{n,\delta,k+1}$ and $\mathcal{P}_{n,{\scriptscriptstyle\mathcal{N}}_{\ns{m}},k+1}$ for $m = 1,2,\ldots,M$ are known as the ``responsibilities'' in EM \cite{bishop2006pattern} and can be expressed as
\begin{equation*}
\begin{aligned}
    \mathcal{P}_{n,\delta,k+1} &= \frac{ \widehat{\rho}_{n,\delta,k}\, \mathcal{CN}\big( \widehat{x}_{n,\textsf{e},k}^+; 0, \gamma_{\bm{x}_{\textsf{e}}^+\ns,k}^{-1} \big) }{ \mathcal{Z}_{n,k+1} },\\
    \mathcal{P}_{n,{\scriptscriptstyle\mathcal{N}}_{\ns{m}},k+1} &= \frac{ \widehat{\rho}_{n,{\scriptscriptstyle\mathcal{N}}_{\ns{m}},k}\, \mathcal{CN}\big( \widehat{x}_{n,\textsf{e},k}^+; \widehat{\mu}_{n,{\scriptscriptstyle\mathcal{N}}_{\ns{m}},k}, \widehat{\gamma}_{n,{\scriptscriptstyle\mathcal{N}}_{\ns{m}},k}^{-1} \ns\ns + \ns \gamma_{\bm{x}_{\textsf{e}}^+,k}^{-1} \big) }{ \mathcal{Z}_{n,k+1} },
\end{aligned}
\end{equation*}

\noindent with the weights $\widehat{\rho}_{n,\delta,k}$ and $\widehat{\rho}_{n,{\scriptscriptstyle\mathcal{N}}_{\ns{m}},k}$, the means $\widehat{\mu}_{n,{\scriptscriptstyle\mathcal{N}}_{\ns{m}},k}$, and the precisions $\widehat{\gamma}_{n,{\scriptscriptstyle\mathcal{N}}_{\ns{m}},k}$ obtained correspondingly from (\ref{eq:BDD-VAMP-EM_pri-xt_parameters}) by substituting $\bm{\theta} = \widehat{\bm{\theta}}_k$. The normalization constant $\mathcal{Z}_{n,k+1}$ is to ensure $\mathcal{P}_{n,\delta,k+1} + \sum_{m=1}^{M} \mathcal{P}_{n,{\scriptscriptstyle\mathcal{N}}_{\ns{m}},k+1} = 1$.

In the M-step, the estimate $\widehat{\bm{\theta}}_{k+1}$ is obtained by maximizing the expected complete-data log-likelihood from (\ref{eq:BDD-VAMP-EM_EM@t_avg-complete-log-lik}):
\begin{equation}
\begin{aligned}
    \widehat{\bm{\theta}}_{k+1} &= [\widehat{\beta}_{\textsf{B},k+1}, \widehat{\beta}_{\textsf{D},k+1}, \widehat{\alpha}_{k+1}, \widehat{\eta}_{k+1}]^\textsf{T}\\
    &\triangleq \arg\max_{\bm{\theta}} Q_{k+1}\big( \bm{\theta}, \widehat{\bm{\theta}}_k \big),
\end{aligned}
\end{equation}

\noindent which decomposes into two independent subproblems:
\begin{subequations}
\label{eq:BDD-VAMP-EM_EM@t_EM-sub-maximizers}
\begin{align}
    [\widehat{\beta}_{\textsf{B},k+1}, \widehat{\beta}_{\textsf{D},k+1}]^{\textsf{T}} &= \arg\max_{[\beta_{\textsf{B}}, \beta_{\textsf{D}}]^{\textsf{T}}} q_{\beta,k+1}(\beta_{\textsf{B}},\beta_{\textsf{D}})\nonumber\\
    &\hspace{0.6cm} \mathrm{subject\,\,to} \quad 
    \begin{cases}
        \beta_{\textsf{B}}\in[0,1]\\
        \beta_{\textsf{D}}\in[0,1]
    \end{cases},\label{eq:BDD-VAMP-EM_EM@t_EM-sub-maximizer1}\\
    [\widehat{\alpha}_{k+1}, \widehat{\eta}_{k+1}]^{\textsf{T}} &= \arg\max_{[\alpha, \eta]^\textsf{T}} q_{(\alpha,\eta),k+1}(\alpha, \eta)\nonumber\\
    &\hspace{0.6cm} \mathrm{subject\,\,to} \quad 
    \begin{cases}
        \alpha\in[0,1]\\
        \eta\in[0,\infty)
    \end{cases}.\label{eq:BDD-VAMP-EM_EM@t_EM-sub-maximizer2}
\end{align}
\end{subequations}

\noindent These maximizations are low-dimensional, i.e., (\ref{eq:BDD-VAMP-EM_EM@t_EM-sub-maximizer1}) is constrained to the unit square $[0,1]^2$, and (\ref{eq:BDD-VAMP-EM_EM@t_EM-sub-maximizer2}) to $[0,1]\times [0,\infty)$. Both can be solved efficiently using standard numerical optimization routines.

\section{Numerical Results}
\label{sec:numerical-results}

This section assesses the performance of BDD-VAMP-EM through numerical simulations with the following primary benchmark algorithms. The sparse Bayesian learning (SBL) algorithm \cite{tipping2001sparse} imposes a parameterized prior to encourage sparsity and utilizes a generalized maximum likelihood (ML) framework \cite{bishop2006pattern} to estimate the hyper-parameters. Benchmarking SBL with BDD-VAMP-EM demonstrates the performance gain achieved by exploring the temporal correlations. The VAMP with EM (EM-VAMP) algorithm \cite{fletcher2017learning} assumes an i.i.d. BG prior for the state vector and jointly performs VAMP signal recovery and EM parameter learning. It serves as the foundational core within BDD-VAMP-EM for the initial time step $t=0$. Benchmarking it allows us to analyze the performance gain attributed to the dynamic BDD model over the static BG prior. The Kalman filter with maximum-likelihood learning (KF-ML) \cite{ghahramani1996parameter} is a standard KF \cite{10.1115/1.3662552} augmented with a ML parameter learning for the auto-regression coefficient and the process-noise variance. This method represents a classic approach for state estimation in linear dynamical systems, which we include to highlight the importance of modeling the sparsity. Moreover, to establish a performance upper bound and evaluate the efficacy of the EM-based parameter learning module, we also include comparisons under an idealized setting in a specific analysis (see Fig.~\ref{fig:plotSyn-SNR}), where the BDD parameters are assumed to be perfectly known. These are the simplified, non-EM variants of the methods above, denoted as VAMP and BDD-VAMP. The AMP-SI \cite{ma2019approximate}, which provides the foundational BDD process model for our work, is also included as a key benchmark.

The evaluation is conducted using synthetically generated data that follows the BDD process for a controlled, parametric analysis. The general system configuration for the simulations is as follows. The BS and UE are equipped with ULAs of $N_{\textsf{Tx}}=64$ and $N_{\textsf{Rx}}=4$ elements, respectively. At each time step, $N_{\textsf{P}}=32$ pilot vectors are transmitted. The pilot symbols are drawn from a quadrature phase-shift keying (QPSK) constellation. Moreover, we set the total number of time steps to $T=100$, the maximum number of internal iterations to $K=1000$, and perform $N_{\textrm{MC}}=50$ Monte-Carlo trials. The performance metric considered throughout the test is the time-averaged normalized root MSE (TNRMSE) between the true signals $\{\bm{x}_t\}_{t=1}^{t=T}$ and their estimates $\{\widehat{\bm{x}}_t\}_{t=1}^{t=T}$, defined as 
\begin{equation*}
    \textrm{TNRMSE}\big(\{\bm{x}_t\}_{t=1}^{t=T}, \{\widehat{\bm{x}}_t\}_{t=1}^{t=T}\big) \triangleq \frac{1}{T} \sum_{t=1}^T \ns{\frac{\left\|\bm{x}_t-\widehat{\bm{x}}_t\right\|_2}{\left\|\bm{x}_t\right\|_2}}.
\end{equation*}
\noindent We also define the sparsity rate $\rho_t \in [0,1]$ as the probability that any given element in $\bm{x}_t$ is zero. The evolution of $\rho_t$ is characterized by the BDD parameters in the following way:
\begin{equation}
\label{eq:syn-data-simulation_rho-evolution}
    \rho_t = (1-\beta_{\textsf{B},t})\, \rho_{t-1} + \beta_{\textsf{D},t}\, (1-\rho_{t-1}).
\end{equation}

\noindent To isolate the impact of specific dynamic properties, we restrict our focus to a time-invariant BDD process. That is, we assume:
\begin{equation}
\label{eq:syn-data-simulation_time-invariant-assumptions}
    \beta_{\textsf{B},t} = \beta_{\textsf{B}},\quad \beta_{\textsf{D},t} = \beta_{\textsf{D}},\quad \alpha_t = 0.9,\quad \eta_t = 1, \quad \forall t.    
\end{equation}

\noindent This assumption yields a steady-state sparsity rate $\rho$, which is obtained by injecting (\ref{eq:syn-data-simulation_rho-evolution}) and (\ref{eq:syn-data-simulation_time-invariant-assumptions}) into $\rho_t = \rho_{t-1} = \rho$ to get
\begin{equation}
\label{eq:syn-data-simulation_rho}
    \rho = \frac{\beta_{\textsf{D}}}{\beta_{\textsf{B}} + \beta_{\textsf{D}}} = \frac{1}{1 + \frac{\beta_{\textsf{B}}}{\beta_{\textsf{D}}}}.    
\end{equation}

\subsection{Performance versus SNR}
\label{subsec:numerical-results_SNR}

\begin{figure}[h!]
\vspace{-0.5cm}
\centerline{\includegraphics[scale=0.38]{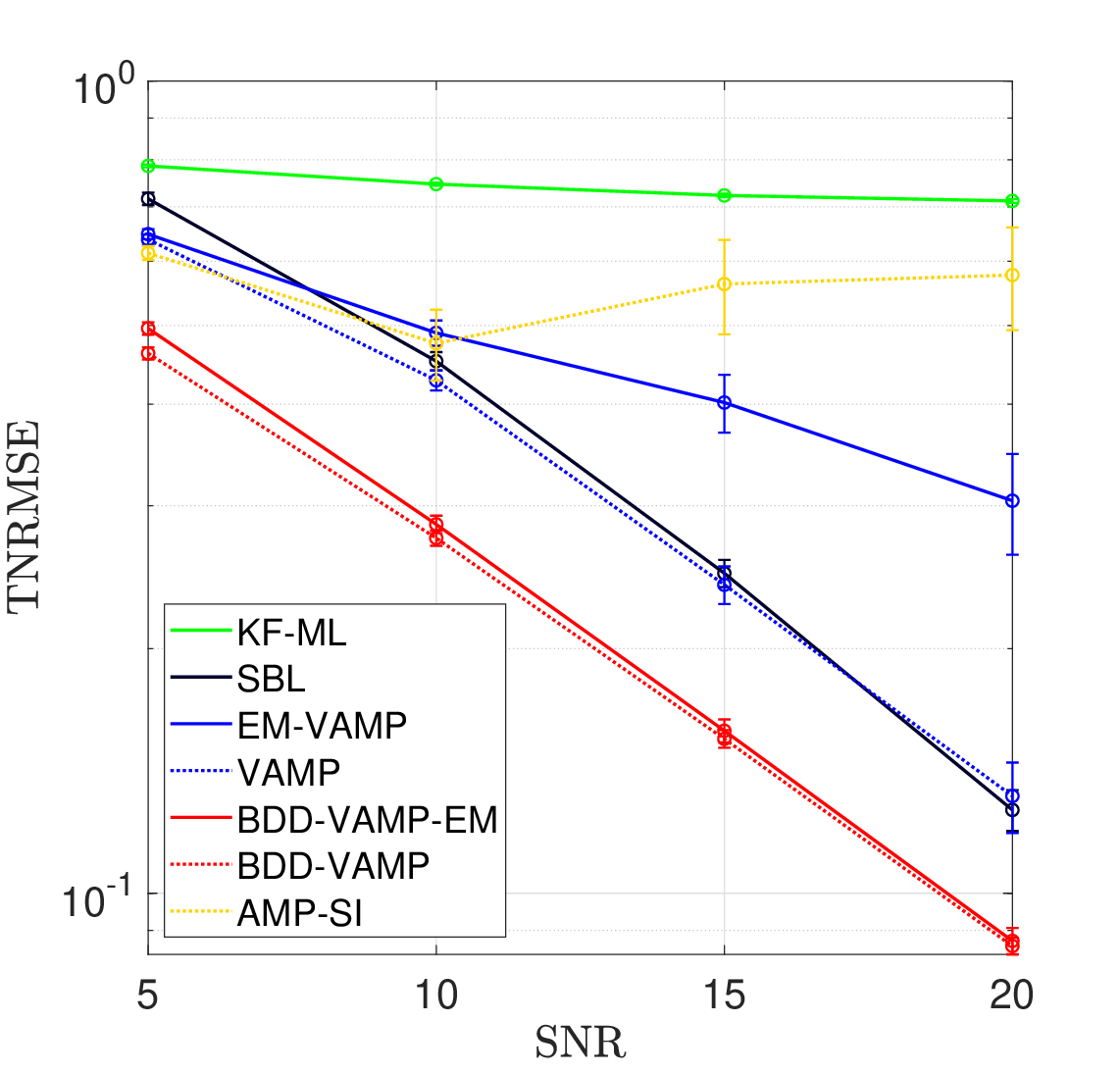}}
    \caption{TNRMSE of BDD-VAMP-EM, KF-ML, SBL, and EM-VAMP, which learn the BDD parameters, and of BDD-VAMP, VAMP, and AMP-SI, which assume perfect knowledge of the BDD parameters, vs. SNR with sparsity rate $\rho=0.8$.}
    \label{fig:plotSyn-SNR}
    \vspace{-0.1cm}
\end{figure}

In Fig.~\ref{fig:plotSyn-SNR}, we investigate how BDD-VAMP-EM performs at SNR levels between 5 [dB] and 20 [dB], with SNR defined as $\textrm{SNR} = \textrm{10 log}_{\textrm{10}}\Big(\frac{ \norm{\bm{H}_t\,\bm{x}_t}^{2}_{2}}{\norm{\boldsymbol{w}_t}^{2}_{2}}\Big)$. We choose $\beta_{\textsf{B}}=0.0125$ and $\beta_{\textsf{D}}=0.05$, which correspond to a steady-state sparsity rate $\rho=0.8$ from (\ref{eq:syn-data-simulation_rho}). The figure shows that BDD-VAMP-EM consistently outperforms all parameter-learning benchmark algorithms at every SNR level. This result justifies the foundational superiority of the BDD model. In comparison to KF-ML, which is based on an auto-regressive model with Gaussian process noise, the BDD model constitutes a more accurate prior for the synthetic virtual channel. Also, in contrast to SBL and EM-VAMP, the BDD model effectively leverages the temporal correlation of the channel, a feature that these benchmarks neglect by performing per-snapshot estimation. Moreover, it is interesting that the sensitivity to SNR varies among the algorithms. BDD-VAMP-EM and SBL exhibit parallel, linear-like convergence with increasing SNR, suggesting a similar asymptotic efficiency. The performance of EM-VAMP plateaus at a slower rate, accompanied by greater variability. Most strikingly, the performance of KF-ML is largely invariant to SNR, which underscores a fundamental limitation of its model in this specific estimation context. Regarding the parameter-known benchmarks, two key observations emerge. First, the performance gap between BDD-VAMP-EM and BDD-VAMP is notably smaller than that between EM-VAMP and VAMP. This indicates the robustness of BDD-VAMP-EM as its EM-based parameter learning incurs a relatively minor penalty compared to the idealized case with perfect knowledge. Second, AMP-SI exhibits a non-monotonic TNRMSE trend with increasing SNR. This counterintuitive behavior arises from the Kronecker-product structure of the effective sensing matrix, which introduces statistical dependencies that violate the i.i.d.-Gaussian assumptions of the state evolution in AMP-based methods. As a result, the algorithm's convergence behavior deviates from the ideal case, with performance degradation at higher SNR, where estimation errors become more sensitive to such structural mismatches.

\subsection{Performance versus both $\beta_{\textsf{B}}$ and $\beta_{\textsf{D}}$}
\label{subsec:numerical-results_beta}

\begin{figure}[h!]
\vspace{-0.5cm}
\centerline{\includegraphics[scale=0.38]{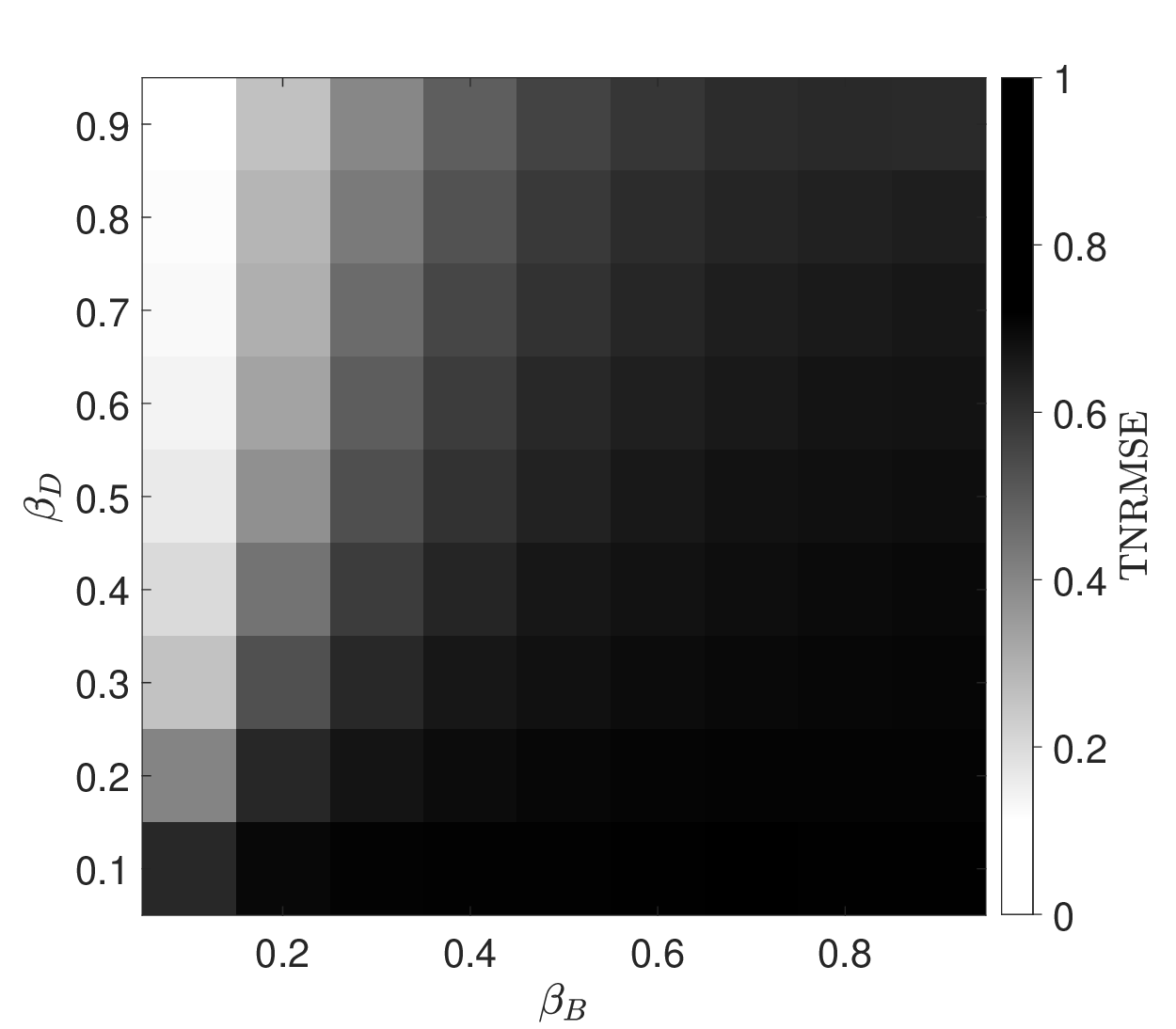}}
    \caption{TNRMSE of BDD-VAMP-EM over a grid of $\beta_{\textsf{B}}$ and $\beta_{\textsf{D}}$ at SNR = 15 [dB].}
    \label{fig:plotSyn-beta}
    \vspace{-0.1cm}
\end{figure}

In Fig.~\ref{fig:plotSyn-beta}, we examine the impact of $\beta_{\textsf{B}}$ and $\beta_{\textsf{D}}$ on the reconstruction performance of BDD-VAMP-EM under $\textrm{SNR}=15$ [dB]. The performance exhibits an opposite dependence on $\beta_{\textsf{B}}$ and $\beta_{\textsf{D}}$. For a fixed $\beta_{\textsf{D}}$, an increase in $\beta_{\textsf{B}}$ degrades performance. In contrast, for a fixed $\beta_{\textsf{B}}$, an increase in $\beta_{\textsf{D}}$ improves it. This behavior is governed by the relationship in (\ref{eq:syn-data-simulation_rho}). For decreasing $\beta_{\textsf{B}}$ or increasing $\beta_{\textsf{D}}$ both lead to a higher $\rho$, which enhances the performance of sparsity-aware algorithms like BDD-VAMP-EM. This relationship can be reformulated as $k_{\rho} \triangleq \frac{\beta_{\textsf{D}}}{\beta_{\textsf{B}}} = \frac{1}{\frac{1}{\rho}-1}$, showing that a specific sparsity rate $\rho$ corresponds to a characteristic ratio $k_{\rho} = \beta_{\textsf{D}} / \beta_{\textsf{B}}$. For instance, $\rho$ values of $0.1$, $0.5$, and $0.9$ correspond to $k_{\rho}$ values of $1/9$, $1$, and $9$, respectively. Consequently, a line radiating about the origin $(0,0)$ in the parameter space with slope $k_{\rho}$ represents a contour of the sparsity rate $\rho$. As this line rotates counterclockwise from the $\beta_{\textsf{B}}$-axis ($k_{\rho}=0$) to the $\beta_{\textsf{D}}$-axis ($k_{\rho} \to \infty$), $\rho$ increases from 0 to 1. The general trend of brightening colors along these radial lines confirms the performance improvement with increasing sparsity.

\section{Conclusion}
\label{sec:conclusion}

In this work, we introduced a novel algorithmic framework, dubbed BDD-VAMP-EM, for robust and adaptive channel tracking in dynamic massive MIMO systems. We did so by unifying the BDD model, the VAMP framework, and an embedded EM procedure in a tightly coupled iterative loop. Crucially, the proposed algorithm propagates a richer, hyper-parametric description of the full posterior distribution across time steps, thereby enabling more accurate temporal fusion. Numerical results demonstrate the effectiveness of BDD-VAMP-EM and its superiority over several state-of-the-art algorithms.

\bibliographystyle{IEEEtran}
\bibliography{IEEEabrv,references}

\vspace{12pt}

\end{document}